# Distinguishing Fact from Fiction:

# Student Traits, Attitudes, and AI Hallucination Detection in Business School Assessment

**Authors**:

Dr Canh Thien Dang, King’s Business School, King’s College London, 30 Aldwych, London WC2B 4BG, United Kingdom, canh.dang@kcl.ac.uk (corresponding author)

Dr An Nguyen, King’s Business School, King’s College London, 30 Aldwych, London WC2B 4BG, United Kingdom, an.nguyen@kcl.ac.uk

Canh Dang is a Lecturer in Economics at King's Business School, King's College London. Dang is also affiliated with the Institute of Finance and Technology, University College of London, and the London School of Economics. Dang publishes in journals such as Economic Development and Cultural Change, Journal of Economic Behavior and Organization, and Oxford Economic Papers. Dang is interested in the effects of AI on education and the economy.

An Nguyen is Lecturer in Human Resource Management at King's Business School, King's College London. Nguyen is affiliated with the University of Glasgow and Open University. Nguyen is interested in using quantitative, qualitative and machine learning techniques to conduct research in HRM, Education Policy and Human Capital Management in global contexts. Nguyen publishes in education-focused and human capital journals. Nguyen is interested in pedagogical research on the use of AI and human capital policies.

**Acknowledgement**

We are grateful for the generous funding from the Innovative Education Fund, King's Business School (KCL), to Dragos Radu for his support in setting up the exercises, to Sally Everett for supporting our idea, and to Alison McFarland, Petar Stankov, and Dirk Lindebaum, and participants at the AEA CTREE (2024), KCL's Festival of AI, and KCL Education Conference; CoP CTaLE AI seminars for insightful comments that help shaped the paper. We use AI tools such as Grammarly and Microsoft Word for checking grammatical errors and spelling. As it will become transparent, we use ChatGPT 3.5 to generate a response to a coursework question, which forms part of the assessment for students to evaluate. We are indebted to an anonymous referee and Antonio Mele, on a separate occasion, for raising the critical question on the long-term significance of AI Hallucinations. Finally, we thank the editor and the referees for key suggestions that improve our theoretical discussion. All remaining errors are ours and we do not represent the views of the funder on this topic.

# Distinguishing Fact from Fiction:

# Student Traits, Attitudes, and AI Hallucination Detection in Business School Assessment

**Abstract**

As artificial intelligence (AI) becomes integral to the society, the ability to critically evaluate AI-generated content is increasingly vital. On the context of management education, we examine how academic skills, cognitive traits, and AI scepticism influence students' ability to detect factually incorrect AI-generated responses (hallucinations) in a high-stakes assessment at a UK business school (n=211, Year 2 economics and management students). We find that only 20% successfully identified the hallucination, with strong academic performance, interpretive skills thinking, writing proficiency, and AI scepticism emerging as key predictors. In contrast, rote knowledge application proved less effective, and gender differences in detection ability were observed. Beyond identifying predictors of AI hallucination detection, we tie the theories of epistemic cognition, cognitive bias, and transfer of learning with new empirical evidence by demonstrating how AI literacy could enhance long-term analytical performance in high-stakes settings. We advocate for an innovative and practical framework for AI-integrated assessments, showing that structured feedback mitigates initial disparities in detection ability. These findings provide actionable insights for educators designing AI-aware curricula that foster critical reasoning, epistemic vigilance, and responsible AI engagement in management education. Our study contributes to the broader discussion on the evolution of knowledge evaluation in AI-enhanced learning environments.

# 1. INTRODUCTION

With the release of Chat Generative Pre-Trained Transformer (ChatGPT) in November 2022, Large Language Models (LLMs) have rapidly transformed educational practices, offering new avenues for student engagement and subject understanding (Fütterer et al., 2023). The proliferation of GenAI tools poses challenges for educators, particularly in designing assessments that maintain academic integrity while embracing generative AI (GenAI) as a learning tool (Moorhouse et al., 2023). A broader societal concern is the tendency of GenAI to generate hallucinations—false yet seemingly coherent information, which, if undetected, can propagate misinformation. This paper examines how students in a UK business school navigate this issue. We assess their ability to detect AI hallucinations in a high-stakes context and then explore how exposure to misinformation influences future performance in a similar higher-stake task. While framed within the context of management education, our paper aligns with broader discussions on misinformation detection, cognitive biases, and epistemic reasoning (Angelucci & Prat, 2024; Peren Arin et al., 2023). By situating students in a setting where they have resources and motivation to evaluate AI-generated responses, our study sheds light on the necessary skillset for critical AI engagement in education. Given the increasing reliance on AI in our society, we contribute to the growing debate on the ethical use of AI (Larson et al., 2024; Moser et al., 2022), and its roles in professional settings (Brynjolfsson et al., 2025). Our findings inform AI literacy scholarship, demonstrating how structured assessment strategies can cultivate critical reasoning, epistemic cognition, and AI scepticism, ensuring that students engage not just as users but as evaluators of AI-generated content.

Using data from a Year 2 econometrics course for economics and management students, we find a significant correlation between academic ability, critical thinking, and the capacity to detect and articulate concerns about AI-generated hallucinations. Despite having access to ample resources, only 20% of the cohort identified hallucinations five weeks into the term, underscoring both the prevalence of AI hallucinations and the need for stronger educational interventions to support students in discerning misinformation.[1] While rote memorisation and knowledge application play a role in learning, they appear less effective in recognising the subtle inaccuracies masked by AI's coherent and eloquent writing. Instead, strong synthesis and evaluation skills emerge as key indicators of hallucination detection. Gender differences are also notable, with female students more likely to detect AI errors, potentially due to their stronger academic performance in the course and a generally more cautious stance towards AI (Stein et al., 2024). We also show that students who identified hallucinations also expressed greater scepticism (more negative) towards AI.

The detection ability, nevertheless, did not translate into a greater likelihood of choosing an optional exam question on the same topic or achieving higher overall performance. Our specific, timely and coursework-related feedback strategy seems to have bridged the gap between students who initially succeed in detecting hallucinations and those who do not. This finding aligns with Transfer of Learning Theory (Perkins & Salomon, 1989), which suggests that structured reinforcement of learning experiences enables students to apply necessary skills in future settings of similar or more advanced nature. Feedback on AI hallucination detection serves as a scaffold for knowledge transfer, allowing students to refine their cognitive strategies and approach similar tasks with increased confidence.

We make three key theoretical contributions to AI literacy, epistemic cognition, and management education theory. First, we extend epistemic cognition theory (Chinn et al., 2011) by empirically examining how students engage in knowledge evaluation and justification in AI-integrated learning environments. Prior research has primarily focused on epistemic cognition in traditional educational settings, but ours applies to the novel challenge of AI hallucination. We demonstrate that students with stronger academic credentials, language proficiency, and epistemic scepticism are more adept at identifying AI-generated misinformation. This contributes to a more nuanced understanding of AI literacy: evaluating AI outputs is not just a technical skill but a cognitive and epistemic process.

[1] In Angelucci & Prat (2024), 50% of the sample are uncertain when choosing between a fake and true story.

Second, we advance cognitive bias and decision-making theories by demonstrating the role of automation bias (Mosier & Skitka, 2018) and anchoring bias (Tversky & Kahneman, 1974) in AI hallucination detection. While previous studies have explored automation bias in professional decision-making contexts, we apply it to learning and AI engagement, showing that students with higher scepticism toward AI are less susceptible to over-reliance on automated outputs. We highlight gender-based differences in AI hallucination detection, with new evidence beyond the established gender gap in trust and usage in technology (Aldasoro et al., 2024), contributing to debates on how cognitive biases shape AI interactions in education.

Third, our study bridges AI literacy with management education theory, contributing to the growing discourse on how AI impacts learning, assessment, and student decision-making. Traditional management education has emphasised critical thinking and analytical skills, while we introduce AI scepticism, epistemic vigilance, and cognitive adaptability as essential competencies for AI-integrated learning environments. It also informs curriculum design, demonstrating that rote memorisation and procedural knowledge are insufficient for AI misinformation detection and that educational programmes must prioritise interpretive reasoning, epistemic reflection, and sensemaking exercises.

By integrating epistemic cognition, cognitive biases, and transfer of learning theory, we provide a theoretically robust framework for understanding how students critically engage with AI-generated content. It challenges the assumption that AI literacy is solely about technical proficiency and instead positions it as an advanced epistemic skill that must be explicitly developed within management education. Our findings have implications for faculty training, curriculum development, and assessment strategies, ensuring that students are prepared not just to use AI but to critically evaluate its outputs in academic and professional decision-making.

Beyond its theoretical contributions, our study offers the first to examine GenAI hallucinations using human evaluators with subject-matter expertise. We propose a flip-classroom assessment approach that lets students to detect and evaluate AI errors while academics evaluate the quality of student's evaluations. Recent research on AI hallucination detection has primarily focused on automated techniques to assess and mitigate errors in AI-generated content, mostly on news summarisation or language translation. Methods such as entropy-based uncertainty estimators have been used to gauge confidence levels in AI outputs (Farquhar et al., 2024), while extraction-based and LLM-driven systems have been tested in various contexts to identify hallucinations in decision-making tasks (ChakrabortyNeeloy et al., 2024). Other frameworks rely on statistical analysis and factual verification to refine hallucination detection (Tam et al., 2023). Our human-centred approach extends beyond automated methods, providing new insights into how domain knowledge, cognitive skills, and epistemic reasoning shape hallucination detection.

Finally, we contribute a practical framework for integrating AI literacy in management education by embedding structured exposure to AI-generated misinformation into assessments and assess how students scrutinise, verify, and contextualise AI-generated content. The approach is scalable to develop epistemic cognition, metacognitive regulation, and AI scepticism among students. We contribute to discussions on AI's role in management education (see Valcea et al., 2024, Larson et al., 2024), ensuring that future business leaders are equipped to navigate AI-driven decision-making responsibly and critically.

## 2. CONCEPTUAL FRAMEWORK

We begin by examining the concept of AI hallucinations before outlining the theoretical rationale for empirical setting and data collection. We then develop a theoretical framework to support our hypotheses, integrating insights from epistemic cognition, cognitive bias, and transfer of learning theories.

### 2.1. What is AI hallucination?

AI hallucinations, as defined by Zhang et al., (2023) and McIntosh et al., (2024), fall into three categories: input-conflicting, context-conflicting, and fact-conflicting. Input-conflicting hallucinations occur when LLMs generate content that contradicts user inputs, producing responses unrelated to the request. Context-conflicting hallucinations stem from inconsistencies within AI-generated text, often due to limitations in maintaining coherence across longer interactions. Fact-conflicting hallucinations involve fabricated or misleading information that contradicts established knowledge, potentially leading to misconceptions. Our paper specifically examines how GenAI produces factually incorrect responses that are inconsistent with the subject knowledge taught to students.

### 2.2. Cognitive load theory: High-stake assessment and hallucination detection.

Our empirical setting, detecting AI hallucinations in a summative assessment component, is founded in the cognitive load theory (Sweller, 1994). The theory distinguishes between (i) intrinsic cognitive load, which depends on task complexity, (ii) extraneous load, which arises from unnecessary distractions, (iii) and germane load, the cognitive effort dedicated to deep learning and problem-solving. Under appropriate settings, students can be appropriately motivated by a high-stakes assessment context to exert cognitive load, with adequate, if not abundant, time to process information to make the best evaluation.

Detecting AI hallucinations is a high-cognitive load task, requiring critical evaluation, epistemic cognition, and structured verification rather than surface-level recognition. Assessing students in a coursework setting rather than a time-constrained exam minimises extraneous cognitive load, allowing them to focus on analysing AI inconsistencies rather than managing time pressure. This setup mirrors real-world decision-making, where professionals must critically assess AI-generated content without strict time pressure. Our setting differs from the current studies using AI-generated content in a non-assessment context (low stakes environment, such as in class content, Hyde et al., 2024) where students are less likely to exert the same cognitive effort to distinguish between facts and hallucinations. By balancing task complexity and cognitive resources, our assessment setting ensures a rigorous but manageable environment, enabling students to demonstrate true AI literacy skills. If the cognitive load were too high, students might rely on heuristics, masking their actual ability to detect hallucinations, while a low-stakes setting might fail to capture the level of scrutiny applied in professional contexts. This approach provides a meaningful test of AI hallucination detection, aligning both cognitive theory and real-world applicability in management education.

### 2.3. Theories behind the predictors of hallucination detection abilities

The extant academic literature on GPT hallucinations predominantly focuses on the technicalities of the algorithms and training data, often marginalising the human factors during the evaluation process (McIntosh et al., 2024). We focus on a more practical and relevant exercise: the ability of students, who were taught and trained with relevant materials, to detect and articulate GPT hallucinations under a high-stake setting with adequate time allowance. Our first research question is formulated as follows:

*Research question 1: What are the predictive factors of students' ability to detect/articulate factually inaccurate information generated by AI?*

The cognitive and epistemic mechanisms that influence the ability to detect AI-generated hallucination are underexplored in management education. While critical thinking and AI scepticism are intuitively thought of as important competencies, a theoretical foundation is needed to understand the specific cognitive and academic factors that influence AI hallucination detection. To address this gap, we develop a conceptual framework grounded in epistemic cognition, cognitive biases, and metacognitive regulation, positioning AI hallucination detection as a function of academic performance, language proficiency, gender differences, and AI scepticism.

Building on the epistemic cognition theory (Chinn et al., 2011), we first conceptualise how systematic cognitive biases shape students' interpretation of AI-generated content, influencing their ability to detect AI hallucinations. Automation bias leads individuals to over-rely on AI outputs, assuming credibility due to the system's perceived intelligence (Mosier & Skitka, 2018). Confirmation bias suggests that students may overlook hallucinations if they align with pre-existing knowledge (Nickerson, 1998); while authority bias implies that AI’s seemingly eloquent, unambiguous, and authoritative tone may discourage scrutiny (Milgram, 1963). Additionally, anchoring bias suggests that students may struggle to override their initial impressions (the anchor) when AI responses mix correct and incorrect information, making it harder to detect hallucinated content (Tversky & Kahneman, 1974).

These biases interact with the epistemic cognition of individuals, which explains how they evaluate, justify, and revise knowledge, particularly when faced with ambiguous or misleading information. In the context of AI-generated content, students must engage in knowledge evaluation by assessing credibility instead of assuming correctness, justification by identifying inconsistencies (articulation), and revision by adjusting their beliefs upon detecting hallucinations (epistemic reflexivity). AI hallucination detection requires strong epistemic awareness as AI-generated content often appears plausible but may be incorrect. The effectiveness of this process depends on three broad sets of key predictors: academic knowledge, language proficiency, and AI scepticism.

A strong academic foundation enhances students’ epistemic cognition—the ability to critically evaluate knowledge, justify beliefs, and revise understanding based on new evidence. Students with stronger current and past academic performance are more likely to possess the necessary analytical reasoning and subject-specific expertise to detect inconsistencies in AI-generated content. This aligns with cognitive load theory (Sweller, 1994), which posits that students with well-developed knowledge structures experience reduced extraneous cognitive load, enabling them to focus on identifying errors in AI-generated text rather than struggling with comprehension. Transfer of learning theory (Perkins & Salomon, 1989) suggests that prior academic success facilitates the application of analytical skills in new contexts, including AI evaluation. These theories collectively support Hypothesis 1a, which predicts academically better students are more likely to detect hallucinated content.

*Hypothesis 1a: Students with better current and past relevant academic credentials are more likely to detect/articulate AI hallucination.*

Linguistic proficiency plays a similarly critical role in AI hallucination detection, as students must not only recognise inconsistencies in AI-generated text but also articulate errors coherently. Epistemic cognition theory suggests that stronger language skills facilitate deeper engagement with textual content, enabling students to identify logical inconsistencies, syntactic ambiguities, and misleading coherence in AI outputs. This is further supported by metacognitive regulation theory (Flavell, 1979), which highlights the importance of self-monitoring and reflective thinking in evaluating complex information. AI-generated text often exhibits fluency illusions—seemingly coherent but incorrect responses that can mislead less proficient readers. Thus, students with stronger writing and interpretive skills are better equipped to scrutinise AI-generated content, assess its credibility, and communicate inaccuracies effectively. This theoretical grounding supports the following Hypothesis 1b:

*Hypothesis 1b: Students with better language skills are more likely to detect/articulate AI hallucination.*

Beyond academic and linguistic skills, cognitive biases and epistemic cognition shape students' interaction with AI-generated content. Research on risk perception and technology trust suggests that women exhibit greater caution toward AI (Aldasoro et al., 2024), aligning with metacognitive regulation theory (Flavell, 1979), which posits that individuals interpret ambiguous information based on prior beliefs and contextual cues. Epistemic cognition research further suggests that women are more likely to engage in reflective assessment of new information, questioning its validity rather than accepting it at face value (see a review on theories using social–psychological, anthropological–evolutionary, and medical science evidence by Meyers-Levy & Loken, 2015). This gender difference, termed "epistemic vigilance" (Sperber et al., 2010), enhances error detection, making female students more likely to scrutinise AI-generated responses for inconsistencies. In contrast, higher technological self-efficacy among men may lead to faster but less critical acceptance of AI outputs. These gender-linked variations in epistemic cognition provide a basis for Hypothesis 1c:

*Hypothesis 1c: Female students are more likely to detect/articulate AI Hallucination.*

AI scepticism is another important cognitive factor influencing hallucination detection. Epistemic vigilance suggests that individuals who are naturally more sceptical engage in higher levels of cognitive scrutiny when encountering new information (Sperber et al., 2010). Students who are predisposed to questioning AI reliability are more likely to verify outputs, identify inconsistencies, and resist automation bias. This is reinforced by metacognitive regulation theory, which highlights how sceptical individuals actively monitor and adjust their beliefs in response to perceived uncertainty in AI-generated content. Additionally,, Pennycook et al., (2020) argue that individuals with greater scepticism toward misinformation are better at distinguishing fact from fabrication, a concept directly applicable to AI hallucination detection. We formulate the link between scepticism and detection ability as follows:

*Hypothesis 1d: Students who are more negative (sceptical) towards AI are more likely to detect/articulate AI Hallucination.*

### 2.4. Detecting AI hallucinations and subsequent performance on related task

We next examine the long-term impact of detecting AI hallucinations on students' academic development and decision-making skills. If detecting hallucinations strengthens critical reasoning and epistemic vigilance, does this effect persist beyond immediate assessments? Does it influence students' broader engagement with AI-generated content in similar settings? Understanding these long-term consequences is essential for integrating AI literacy into management education as a sustainable and long-term competency rather than a one-time skill.

The link between AI hallucination detection and future performance in high-stakes tasks can be explained through transfer of learning theory (Perkins & Salomon, 1989) and Epistemic Cognition (Chinn et al., 2011). Transfer of learning suggests that cognitive skills developed in one context—such as critically evaluating AI-generated content—enhance problem-solving in future analytical tasks, particularly in structured assessments requiring similar reasoning. Students who successfully detect hallucinations may develop stronger self-efficacy, a key component of self-regulated learning (Zimmerman, 2002) which increases confidence and engagement in similar high-stakes tasks, such as final exams. When mastery experiences—like correctly identifying AI hallucinations—are reinforced through specific, timely feedback students are more likely to refine their cognitive strategies and improve future performance (Shute, 2008). Conversely, students who fail to detect hallucinations but receive constructive feedback may either improve their cognitive strategies or experience reduced confidence in handling similar content in a similar task in the future (Fedor et al., 2001). Epistemic cognition further supports this link, as students who critically evaluate AI-generated information develop habits of inquiry, justification, and revision, strengthening their ability to assess credibility and solve problems in later assessments. Alternatively, our timely and targeted feedback may bridge the gap from the initial ability to engage with the material examined in the AI-related question, resulting in an equal performance in the final exams. Empirically, we examine another high-stake setting where students choose whether to engage with a similar AI-related question to the one they previously evaluated. We hypothesise:

*Hypothesis 2: Students who successfully detect AI hallucinations are more likely to engage in and perform well in similar tasks in future high-stakes settings (e.g., final exams).*

We summarise the theories behind our framework that links our high-stake environment, predictors of AI hallucination detection, and its potential long-term impact in Table 1, with a visual aid in Figure 1.

**Table 1.** A summary of key established theories underpinning the empirical setting, and hypotheses

| Theoretical Lens | Key Concepts | Application to AI Hallucination Detection | Hypothesis |
|---|---|---|---|
| Epistemic Cognition Theory (Chinn et al., 2011) | Critical evaluation, justification, and revision of knowledge | Students who engage in structured knowledge evaluation and challenge AI-generated misinformation develop stronger reasoning skills, improving future performance. | H1a & H1b; H2 |
| Cognitive Load & Transfer of Learning (Sweller, 1994; Perkins & Salomon, 1992) | Prior knowledge reduces cognitive strain and improves application of skills | Students with stronger foundations experience lower cognitive load, easier to detect subtle inconsistencies, and transfer evaluation skills to future high-stake assessments. | H1a; H2 |
| Metacognitive Regulation (Flavell, 1979) | Interpreting ambiguous information and self-monitoring | Students who identify AI errors develop self-monitoring strategies, leading to better reasoning and verification in future assessments. | H1b |
| Automation Bias & Epistemic Vigilance (Mosier & Skitka, 2018; Sperber et al., 2010) | Tendency to over-rely on AI vs. scepticism-driven verification | Students with higher scepticism toward AI are more likely to question outputs, fact-check, and identify hallucinations over time | H1d; H2 |
| Gender Differences in Risk Perception & AI Trust (Aldasoro et al., 2024) | Differences in technology scepticism and decision-making | Female students exhibit greater AI scepticism and risk awareness, making them more likely to detect hallucinations. | H1c |

*Notes:* The Cognitive Load Theory (Sweller, 1994) underpins our choice of a high-stake but take-home assessment in which adequate time and high stakes ensure students allocate cognitive resources to deep analysis and verification rather than relying on heuristics or being overwhelmed by time pressure, providing a realistic measure of AI literacy in management education.

*Figure 1.* A flowchart of hypothesis development, key theoretical linkage (in bold), and the data source

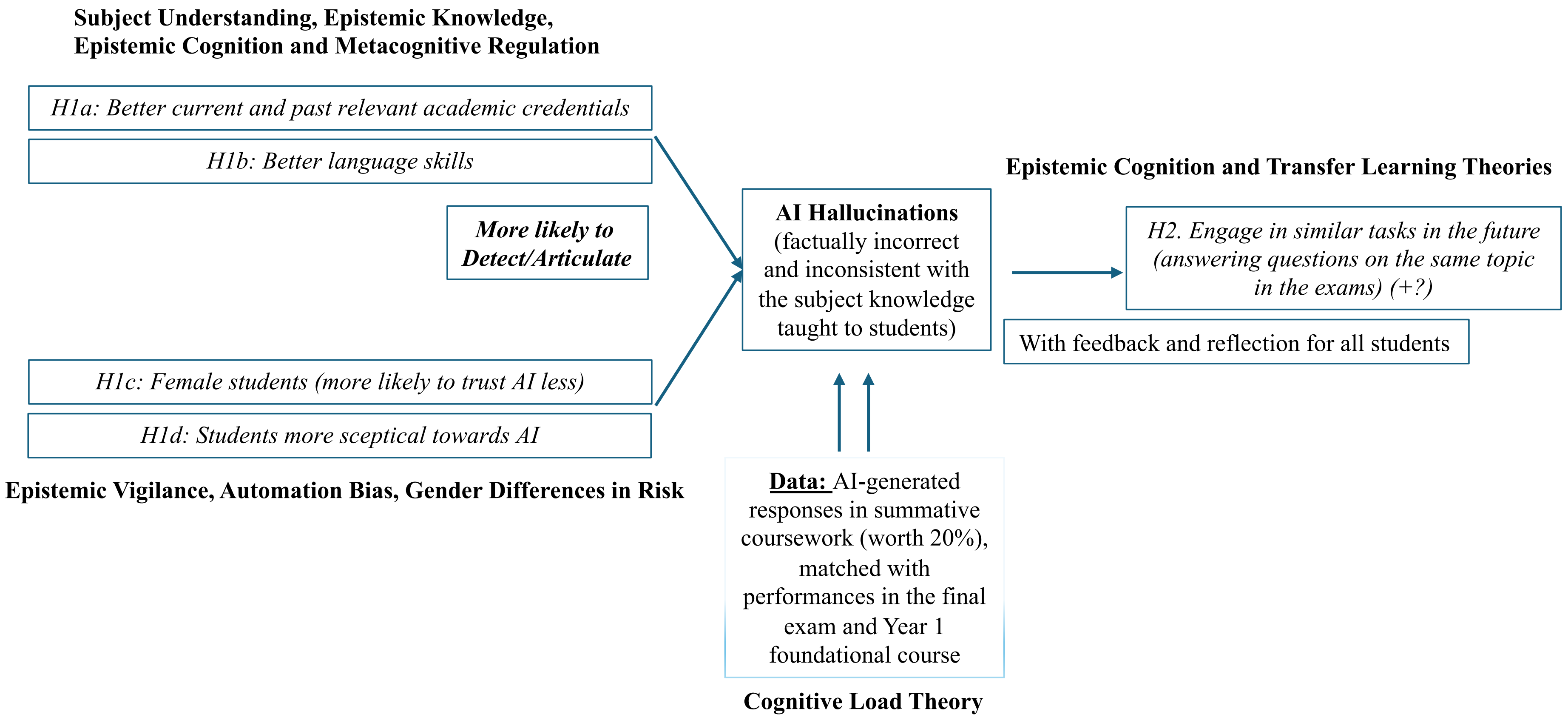

## 3. RESEARCH DESIGN

We discuss how the data is collected and the context of the empirical analysis, followed by a discussion of how we measure different variables of interest.

### 3.1. The Year 2 Econometrics Course in our data

Our data come from a Year 2 economics coursework assessment at a UK business research-led school. UK economics degrees typically span three years: Year 1 covers introductory economics, statistics, and mathematics; Year 2 includes intermediate economics and econometrics; and Year 3 offers specialised optional courses. Our econometrics course, compulsory for BSc Economics and BSc Economics & Management students, builds on Year 1 *Statistics for Economists* and covers linear regression, inference, panel data, instrumental variables, and difference-in-difference methods over 10 weeks. Assessment includes a coursework (20% of the grade) in Week 5 (end of October, marks and detailed and personalised feedback were provided within three weeks of submission) and an online exam (80%) in January. We analyse student performance from both assessments, with detailed sub-question marks, for 211 students.

### 3.2. The coursework and the exams

By the time of the coursework, students have covered linear and multiple regression analysis, including omitted variable bias, selection bias (with randomized control trial examples), inference, and statistical interpretation. We identify from early on that *selection bias* is a topic that confuses the students. The definition of selection bias remains controversial in epidemiology and existing definitions of the term are often ambiguous (Lu et al., 2022). We carefully and repeatedly discuss the concept of *selection bias* that is standard in causal study and econometrics as the *"inherent difference between the treatment and the control group even* *in the absence of the treatment",* which is a threat to internal validity.[2] It is in contrast a different definition of selection bias that focuses on "the non-representativeness of the sample to the population of interest when the sampling is not random", which we insist on calling *sample selection bias* to distinguish from the other concept. While sample selection bias is introduced in Year 1 Statistics, we reinforce these concepts through a revision lecture to ensure students are well prepared. To assess their understanding, we include Question 1 in the coursework, worth 15 out of 100 marks, focusing on basic statistical testing and data interpretation, reflecting key learning outcomes from their prior statistics training. Figure 2 presents the GenAI response to our econometric question and the evaluation criteria used to assess students' comprehension.

We design Questions 2 and 3 to assess different cognitive skills. Question 2 requires students to interpret results using econometrics knowledge, particularly selection bias, which relates to the AI question. Question 3 is procedural, asking students to follow instructions and perform statistical tests. This distinction allows us to examine interpretive reasoning in Question 2 and procedural knowledge in Question 3. For the AI-related task, students receive 10 out of 100 marks for detecting AI hallucinations, with instructions to assess only the econometrics content, not the writing quality of AI responses. A follow-up question requires a 100-word reflection on AI, serving two purposes: (1) capturing students' immediate sentiment towards GenAI, and (2) measuring English writing proficiency, providing a more objective assessment than self-reported fluency. The two AI questions contribute 15% of the coursework grade. Marks, solutions, and detailed feedback are released within term time, ensuring students understand the AI errors before the final exam. Students, by the final exam, are aware of the AI hallucinations and the correct solution to the question.

---

[2] It is often defined as one special case of "collider stratification bias" in some causality branch of statistics. We do not wish to stray into this terminological discussion in this paper.

***Figure 2.*** The AI question in the mid-term coursework (worth 20% of the total grade).

6. **A question using information generated from a Generative AI**

   i. **(10 marks)** You can find on the next page an econometrics question and the answer generated by ChatGPT 3.5. *Concisely* and *critically* evaluate this answer **using the econometrics knowledge you have learned so far**. [*Hints*: You can evaluate the answer point by point, paying attention only to its econometric content, not the quality of writing or cohesion.]

   ii. **(5 marks)** Provide a short reflective paragraph (up to 100 words) about what you learned from this exercise. [*Hints*: There is no right or wrong approach to answering (ii.)]

***The following question is inputted to ChatGPT 3.5 on 15 August 2023.***

> Explain the concept of selection bias in econometrics and provide an example to illustrate this concept. Discuss the potential consequences of selection bias on the validity of estimation results in regression analysis.

***Below is the response from ChatGPT:***

> **Selection Bias in Econometrics:**
> Selection bias refers to a situation in econometric analysis where the sample used for estimation is not representative of the entire population due to a non-random selection process. This bias can lead to distorted estimation results and misleading inferences about relationships between variables.
>
> **Example of Selection Bias:**
> Suppose an economist is examining the relationship between education and income using survey data that only includes individuals who are currently employed. This could introduce selection bias, as those with lower education levels who are not currently employed are excluded from the analysis. This bias might lead to an overestimation of the positive impact of education on income.
>
> **Consequences of Selection Bias:** Selection bias can have several negative consequences on the validity of estimation results in regression analysis:
>
> 1. Inaccurate Coefficient Estimates: Selection bias can lead to coefficient estimates that are biased and do not accurately reflect the true relationships between variables. The estimated effect of the variable of interest becomes confounded by the selection process.
>
> 2. Inefficient Estimation: Biased estimates resulting from selection bias are also likely to be inefficient, leading to wider standard errors. This affects the precision of parameter estimates and the ability to detect true relationships.
>
> 3. Incorrect Inference: If selection bias is not appropriately accounted for, the estimated relationships may not be generalizable to the broader population, leading to incorrect policy recommendations or business decisions.
>
> 4. Invalid Hypothesis Testing: Selection bias can invalidate hypothesis tests, leading to incorrect conclusions about the statistical significance of relationships.

The final exam, worth 80% of the course, takes place two months later and includes an optional short-answer section, where students choose four out of six questions (each worth 5%). We deliberately include a sub-question on selection bias, directly linked to the AI coursework. By matching coursework and exam performance, we assess whether students choose to answer this question after critically engaging with AI-generated errors, addressing Research Question 2.

## 3.3. The marking

Markers are instructed to evaluate responses based on econometric content, not writing quality (except for Question 6ii on the reflective paragraph). For the AI hallucination question, students must correctly identify the AI's misclassification of selection bias as sample selection bias to score above 5 marks (40% pass threshold), ensuring they recognize factual inconsistencies. This marking scheme distinguishes those who detect hallucinations from those who do not and evaluates how well students articulate the correct response. The reflective paragraph is marked more flexibly, rewarding critical thinking rather than correctness.

## 3.4. Variable Description

### *3.4.1. Outcome variables*

We focus on two set key outcome measures. First, we define a binary variable, *Detect*, which equals 1 if a student correctly identifies factually incorrect and inconsistent information in the AI response, and 0 otherwise. Related, we use *AI Mark*, which captures the extent to which students critically evaluate the AI response, as reflected in their awarded marks for the question.

Second, to examine the relationship between AI hallucination detection in coursework and engagement in a higher-stakes setting (final exam), we define *Answer AI in Exam*, a binary variable equal to 1 if a student chooses to answer the optional exam question related to the AI coursework question, and 0 otherwise. Additionally, we use *Exam AI Mark* to measure how well students perform on this related exam question, assessing their ability to apply prior learning in a high-stakes context.

### *3.4.2. Predictors of AI hallucination Detection*

Given the data availability, we are interested in three broad predictors of students' ability to detect AI hallucinations. (1) students' subject-related academic ability: current and past, (2) language skills and AI sentiment, and (3) gender.

*a) Academic ability*

We obtain three sources of a student's academic ability, (i) *non-AI CW Mark:* the total marks of other questions in the coursework than the AI-response question, (ii) *Statistics Background:* the mark for the coursework's question on Year 1 Statistics materials, (iii) *Exam Mark*: the total mark obtained in the final exam, (iv) *Statistics Year 1 Mark:* the overall mark from the prerequisite Year 1 Statistics course. Because not all students who took Year 1 Statistics progressed to Year 2 Econometrics, we have *Statistics Year 1 Mark* for only 176 students. To identify the key subject skills predictive of a student's ability to detect AI hallucinations, we use the performance for the interpretive tasks (Question 2), *Interpretive Skills,* and for the procedural skills (Question 3), *Procedural Skills*. These two measures directly relate to two cognitive domain levels of Bloom's (1956) knowledge-based taxonomy of students' understanding of a subject: Evaluation and Synthesis for the interpretive skill questions, and Application for the procedural skills questions. Despite various criticisms, Evaluation and Synthesis are generally considered higher-order cognitive process than Application (Woolfolk, 2001). Accordingly, we consider *Interpretive skills* to "better" reflect subject knowledge and understanding compared to *Procedural Skills*. The former relates more closely to the concept of "critical thinking", while the latter addresses a student's attention to detail.

*b) English/Writing Skills and AI sentiment*

We avoid self-reported data, and generic measures due to the subject-specific academic English required in the course. Instead, as part of the AI-related question (Q6ii), students submit a 100-word reflective paragraph on their views on GenAI usage and we obtain two measures of students' writing/English skills and sentiment towards AI.

First, we measure writing skills using the readability of students' 100-word reflective paragraphs, a widely used concept in natural language and semantic research. Readability captures how easily a reader can understand a text, reflecting both language proficiency and cognitive processing. We adopt two established readability metrics: the Flesch-Kincaid Readability Score (*Readability_FKR*, Flesch & Kincaid, 1975), which evaluates sentence length and syllabic complexity, and the New Dale-Chall Readability Formula (*Readability_CDale*, Chall & Dale, 1995), which accounts for semantic familiarity and word difficulty. The two measures are useful in various settings and adopted in leading economics, finance, and management research journals, such as to evaluate (i) textbook readability in economics education (Tinkler and Woods, 2013), and in management education (Loveland et al., 2017); (ii) analyst sell side reports in accounting research ((De Franco et al., 2015)); (iii) annual reports in finance (Hwang & Kim, 2017), and accounting disclosure (Bloomfield, Heinle and Timmermans, 2024); (iv) academic research writing in top economics journals (Rodriguez et al., 2024) and top management journals, (v) product languages (Hoberg and Phillips, 2017); (vi) users' reviews (Shen et al., 2024) in management science. We relegate the full calculations in Appendix A, with technical review in Collins-Thompson (2014).

These metrics are well-suited for assessing writing proficiency in an AI-context assessment, as they capture language processing, syntactic complexity, and argument articulation. Writing complexity serves as a proxy for critical thinking and reasoning skills, both of which are essential for engaging with AI-generated content. In the context of AI hallucination detection, effective writing reflects not only a student's ability to identify errors but also to justify and explain inconsistencies clearly. Unlike self-reported language proficiency, which can be unreliable, these readability measures offer an objective and robust evaluation of students' writing complexity. For *Readability_FKR*, a higher score means a more readable text (easily understood by a younger average person); while for *Readability_CDale*, a lower score corresponds to a more readable text regarding lexical and syntactic complexity.

To enhance the robustness of our writing skills measures, we complement the readability scores with *Marks for Writing Quality*, the score given to Question 6(ii) by our markers, assessing the clarity, cohesion, and readability of students' reflective paragraphs. Unlike algorithmic readability formulas, this human-assigned score captures higher-order writing elements, such as logical structure, argument articulation, and flow, which are critical in evaluating complex AI-generated content. By incorporating assessments from experts in both writing and econometrics, we ensure that the measure reflects not just linguistic complexity but also the ability to formulate coherent and justified reasoning—a key aspect of critical thinking in AI-context assessments. Since readability is influenced by both the writer and the reader, we control for markers' fixed effects and students' academic characteristics to isolate the relationship between writing proficiency and AI hallucination detection. This ensures our findings reflect students' actual language and reasoning abilities, rather than grading variations or external biases. By integrating objective readability metrics with human evaluation, our approach provides a comprehensive assessment of how writing proficiency contributes to students' critical engagement with AI-generated misinformation.

Finally, to assess students' sentiment towards GenAI, we construct an *AI Sentiment* score based on their reflective paragraphs. We adopt the canonical definition of sentiment/attitudes by Ajzen (1991) of an individual's favourable or unfavourable evaluation towards an issue. Assessing students' reflective writings to tease out sentiment is challenging, as traditional evaluation methods—such as rubrics, questionnaires, and thematic coding—often lack depth, validity, or scalability. While qualitative analysis provides richer insights, it requires significant resources, especially for large cohorts. Recent advancements in natural language processing (NLP) and sentiment analysis offer a scalable alternative for analysing student reflections, allowing for the systematic extraction of underlying attitudes, themes, and cognitive engagement with AI-generated content. We rely on computational linguistics and machine learning to classify text as positive, negative, or neutral, regarding how students perceive their learning and dealing with GenAI experiences. This approach is widely used in the management fields (Sheng et al., 2019), or medical education (Zimmerman, 2002). In the context of AI literacy, our approach of analysing students' sentiments toward AI—without directly prompting for self-reported opinions—provides quantifiable insights into their engagement, scepticism, and trust in AI-generated outputs. We obtain the sentiment measures for each student's reflective paragraph using the *Syuzhet* R-package (Jockers, 2023). The method evaluates emotional valence by assigning a score based on predefined word lexicons specifically tailored for words in fiction and literacy studies, which is highly suitable to our reflective paragraphs on students' views towards AI. A higher positive score indicates stronger positive sentiment, while a lower or negative score reflects stronger negative sentiment, with values near zero suggesting neutrality. As a preview of Table 2, our students are largely positive about GenAI with a mean score of 3.47 (min = -8.7).

*c) Gender*

Due to confidentiality, a research assistant anonymises students' names and matches with student IDs of the assessments. Without any other personal information, we estimate students' gender based on their names using algorithm from *genderize.io*. The popular and credited algorithm assigns a probability of gender (female and male) to a first name, using a dataset of more than a billion people's names and genders (Santamaría & Mihaljević, 2018, for a review).

For example, Alexandros and Nicole have a 100% probability of being a male and female name, respectively, while Zhanyu is a male name only with a 76% probability. We take the gender given by the algorithm with a probability of at least 95%. Ambiguous cases are manually verified via LinkedIn or other social media separately by the research assistant.

*d) Markers' fixed effects and Standard errors*

Our marking team consists of six markers for the coursework and five for the exam. Despite standardized grading procedures to ensure consistency, subjective variations may persist. To address this, we include categorical variables, *MarkersCW* and *MarkersEX*, to control for marker fixed effects when analysing coursework and exam data, respectively. Additionally, we account for potential heteroskedasticity by applying robust standard errors throughout our analysis. All hypothesis tests are conducted using two-tailed null hypotheses.

# 4. ANALYSIS AND RESULTS

## 4.1. Summary Statistics

Table 2 presents the summary statistics. Among the 211 students, only 20% correctly identified the incorrect content generated by GenAI, recognising the sample selection bias rather than selection bias. Once students detected the error, they articulated it well, scoring an average of 71%, compared to 41% for non-detectors. Despite this gap, we find no significant unconditional differences between detectors and non-detectors in academic performance, readability skills, or AI sentiment. A large majority (82%) attempted the similar exam question, likely due to prior coursework exposure. The difference here between detectors and non-detectors was not statistically significant. Among the 174 students who answered, the average score was 3.9 out of 5 (nearly 80%), indicating strong performance. Female students were more likely to detect AI hallucinations, as were those who received higher marks for cohesion and writing skills in their reflective paragraphs. Due to exam deferrals, we have 205 observations for Exam Mark, and Year 1 Statistics performance is available for 176 students.

**Table 2.** Summary Statistics for the probability of detecting AI hallucinations and students' characteristics

| | | | | | | ***Detect = 0*** | ***Detect = 1*** | | |
|---|---|---|---|---|---|---|---|---|---|
| | N | Mean | SD | Min | Max | ***Mean*** | ***Mean*** | | t-stat |
| **Variables** | (1) | (2) | (3) | (4) | (5) | *(6)* | *(7)* | (6)-(7) | (8) |
| *Detect* | 211 | 0.199 | 0.400 | 0 | 1 | | | | |
| *AI Mark [10]* | 211 | 4.682 | 1.511 | 0 | 9 | 4.089 | 7.071 | -2.983*** | (-18.62) |
| *Answer AI in Exam* | 211 | 0.820 | 0.385 | 0 | 1 | 0.805 | 0.880 | -0.0762 | (-1.15) |
| *Exam AI Mark [5]* | 174 | 3.862 | 0.830 | 0 | 5 | 3.894 | 3.743 | 0.151 | (0.98) |
| *Non-AI CW Mark [85]* | 211 | 64.592 | 8.872 | 31 | 83 | 64.009 | 66.940 | -2.932 | (-1.93) |
| *Statistics Background [30]* | 211 | 22.754 | 4.078 | 9 | 29 | 22.595 | 23.393 | -0.798 | (-1.14) |
| *Statistics Year 1 Mark [100]* | 176 | 57.705 | 15.447 | 0 | 90 | 56.708 | 62.188 | -5.479 | (-1.83) |
| *Exam Mark [100]* | 205 | 64.295 | 11.952 | 10 | 85 | 64.006 | 65.451 | -1.445 | (-0.69) |
| *Interpretive Skills [15]* | 211 | 6.948 | 1.792 | 2.5 | 11 | 6.873 | 7.250 | -0.377 | (-1.22) |
| *Procedural Skills [15]* | 211 | 11.367 | 2.677 | 1 | 15 | 11.512 | 10.786 | 0.726 | (1.58) |
| *Readability_FKR (Lexical Complexity)* | 211 | 15.060 | 3.469 | -2.62 | 29.415 | 14.942 | 15.536 | -0.594 | (-0.99) |
| *Readability_CDale (Lexical & Syntactic Complexity)* | 211 | 10.808 | 1.127 | 0.1488 | 13.293 | 10.737 | 11.092 | -0.355 | (-1.84) |
| *Marks for Writing [5]* | 211 | 3.642 | 0.769 | 0 | 5 | 3.541 | 4.048 | -0.506*** | (-3.95) |
| *AI Sentiment* | 211 | 3.466 | 4.592 | -8.7 | 16.8 | 3.710 | 2.482 | 1.228 | (1.56) |
| *Gender (=0 if Female)* | 211 | 0.635 | 0.483 | 0 | 1 | 0.680 | 0.452 | 0.228** | (2.79) |

*Notes:* Numbers in brackets [] indicate the maximum possible marks for the questions, and numbers in parentheses () in (8) indicate the t-stat.

## 4.2. Predictors of AI Hallucination Detection and Articulation

We first examine the factors influencing students' ability to detect incorrect and inconsistent AI-generated responses. For interpretability, we report OLS results, as they yield similar qualitative and quantitative findings to Probit/Logit models. Since the only danger of endogeneity to our causal claim is omitted variable bias, we minimise its sources by accounting for two main potential cofounding factors: students' diligence and abilities via our predictors set, and the markers' fixed effects. Nonetheless, our primary objective remains identifying key factors conducive to students' ability to detect AI hallucination. Our estimation is as follows:

$$Detect_i = \alpha + Predictors_i\beta' + MarkersCW_i\gamma_1' + MarkersEX\gamma_2' + e_i \qquad (1)$$

where $i$ indicates student observation, $Predictors_i$ is the set of various potential predictors, $e_i$ is the error terms. $\beta$ is the vector of coefficients for the predictor variables, and $\gamma_1, \gamma_2$ are the vectors of coefficients for the markers fixed effects. Figure 3A presents the estimation results for the probability of detecting AI hallucinations, with full details in Table 2. Figure 3B and the last columns of Table 3 report results using *AI Mark*—the score awarded for evaluating the AI response—replacing *Detect* in Equation (1). This allows us to assess whether the same predictors influence both AI error detection and the ability to articulate critical evaluations. To preview, our findings indicate that similar factors drive both detection and evaluation skills, though the statistical significance is often stronger for the latter.

***Table 3.*** OLS Estimates for predictors of AI hallucination detection and articulation.

| Predictors | **Outcome variable:** | | | *Detect* | | | | *AI Mark* | |
|---|---|---|---|---|---|---|---|---|---|
| | (1) | (2) | (3) | (4) | (5) | (6) | (7) | (8) | (9) |
| *Non-AI CW Mark [85]* | 0.006** (0.003) | | | | | | | | |
| *Statistics Background [30]* | | | 0.016** (0.007) | 0.020*** (0.007) | 0.013* (0.007) | 0.015** (0.007) | 0.015** (0.007) | 0.045* (0.024) | 0.049** (0.024) |
| *Statistics Year 1 Mark [100]* | | 0.003* (0.002) | | | | | | | |
| *Exam Mark [100]* | | | | 0.001 (0.002) | | | | | |
| *Interpretive Skills [15]* | | 0.016 (0.015) | 0.022 (0.015) | 0.021 (0.015) | 0.019 (0.014) | 0.021* (0.012) | 0.021* (0.012) | 0.096* (0.049) | 0.103** (0.050) |
| *Procedural Skills [15]* | | -0.038*** (0.013) | -0.031*** (0.011) | -0.026** (0.012) | -0.03*** (0.011) | -0.03*** (0.011) | -0.03** (0.011) | -0.046 (0.036) | -0.041 (0.037) |
| *Lexical & Syntactic Complexity* | | | | | 0.038* (0.021) | | | 0.141** (0.066) | |
| *Lexical Complexity* | | | | | | 0.000 (0.009) | | | |
| *AI Sentiment* | | -0.012* (0.007) | -0.010* (0.006) | | | | -0.009 (0.006) | | -0.034* (0.018) |
| *Marks for Writing [5]* | | 0.115*** (0.036) | 0.101*** (0.029) | | 0.09*** (0.029) | 0.09*** (0.036) | 0.095*** (0.029) | 0.814*** (0.138) | 0.826*** (0.135) |
| *Gender (=0 if Female)* | | | | -0.094* (0.055) | -0.100* (0.053) | -0.095* (0.053) | -0.097* (0.058) | -0.306* (0.168) | -0.254 (0.172) |
| Markers CW | Yes | Yes | Yes | Yes | Yes | Yes | Yes | Yes | Yes |
| Markers EX | No | No | No | Yes | No | No | No | No | No |
| Observations | 211 | 176 | 211 | 205 | 211 | 211 | 211 | 211 | 211 |
| Adjusted R-squared | 0.109 | 0.168 | 0.193 | 0.166 | 0.202 | 0.190 | 0.200 | 0.396 | 0.395 |

*Notes:* *** p<0.01, ** p<0.05, * p<0.1. Numbers in brackets [] indicate the maximum possible marks for the questions. Robust standard errors are in parentheses. The outcome variable for (1)-(7) is *Detect (*whether a student can detect AI hallucinations), and for (8)-(9) is *AI Mark,* how well the student can articulate their evaluation of the AI-generated response. Fewer observations in (2) and (4). Coefficients for constants and the fixed effects are omitted to save space. We reported only the results using *Statistics Background* as the regressor for academic skills instead of *non-AI coursework* and *Statistics Year 1 Mark* to avoid multicollinearity in columns (2) to (9). The results remain qualitatively similar (statistical significance) if we used any of the academic skill variables.

First, students with **strong academic performance** are more likely to detect and articulate errors in AI-generated responses, as reflected in the positive relationship between the concurrent performance in the non-AI questions in the coursework *non-AI CW Mark* and hallucination detection (Column 1). This supports cognitive load theory, which suggests that well-developed knowledge structures reduce cognitive strain, allowing for more effective analytical reasoning rather than reliance on heuristics. Strong academic engagement fosters critical evaluation skills**,** which are essential for identifying inconsistencies in AI-generated content**.** Breaking down the different academic skills, we find that statistical proficiency—measured by *Statistics Background* and *Statistics Year 1 Mark*—significantly enhances students' ability to detect and articulate AI errors (Columns 2–9). This aligns with epistemic cognition theory, which emphasises the role of domain knowledge and structured reasoning in assessing credibility and refining understanding. A strong foundation in statistics enables students to identify subtle anomalies in AI-generated responses that may go unnoticed by those with weaker subject knowledge. Given data limitations, we prioritise *Statistics Background* in subsequent analysis. These findings highlight the importance of domain-specific expertise in AI hallucination detection, reinforcing the need for curriculum designs that integrate AI literacy within rigorous subject-based learning.

We next examine how interpretive and procedural skills influence students' ability to detect AI hallucinations. Interpretive skills, which involve evaluating and synthesising information, are positively associated with AI error detection, as students who engage in critical analysis and reasoning are more successful in identifying inconsistencies. This aligns with epistemic cognition theory, which emphasises structured knowledge evaluation in detecting misinformation. However, while the relationship is consistent across the specifications, it is less statistically significant, suggesting that interpretive skills play a role but may interact with other cognitive factors. In contrast, procedural knowledge, which focuses on following set methodologies and applying memorised rules, appears to have limited influence on hallucination detection. In fact, students who perform well in procedural tasks are less likely to detect incorrect AI-generated responses, even after controlling for academic performance indicators. This finding suggests that rote application of knowledge alone is insufficient for identifying subtle logical and factual inconsistencies in AI-generated content. Cognitive flexibility, critical reasoning, and credibility assessment are more essential for hallucination detection than simply executing predefined analytical steps. These results reinforce the importance of higher-order thinking skills in AI literacy, highlighting the need to integrate critical evaluation exercises into management education to prepare students for AI-assisted decision-making.

We examine whether overall course performance, measured by *Exam Mark*, predicts students' ability to detect AI hallucinations. Results in Column (4) do not support this hypothesis, even after controlling for academic performance indicators. While *Exam Mark* is strongly correlated with other academic measures, estimations using it as the sole academic predictor yield similar findings, suggesting it does not independently explain hallucination detection. A likely explanation is that detailed feedback and coursework solutions have bridged the performance gap between students who initially detected AI hallucinations and those who did not. As discussed in Section 4.3, the structured learning process appears to have equalised subject comprehension over time, reinforcing the role of targeted feedback in mitigating disparities in AI-critical reasoning skills.

***Figure 3.*** Coefficient plots for the predictors of AI hallucination detection (A) and articulation (B

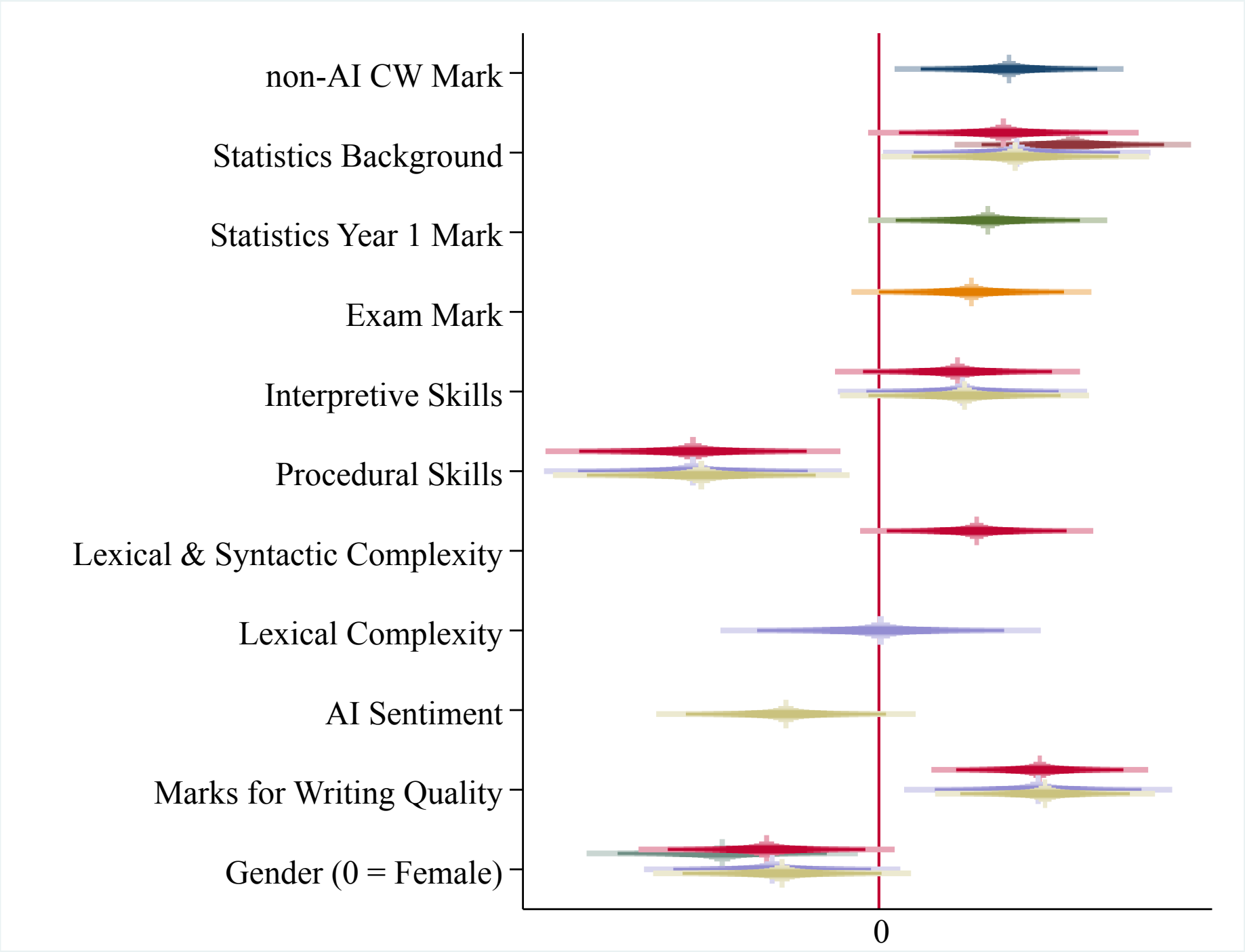

Panel A

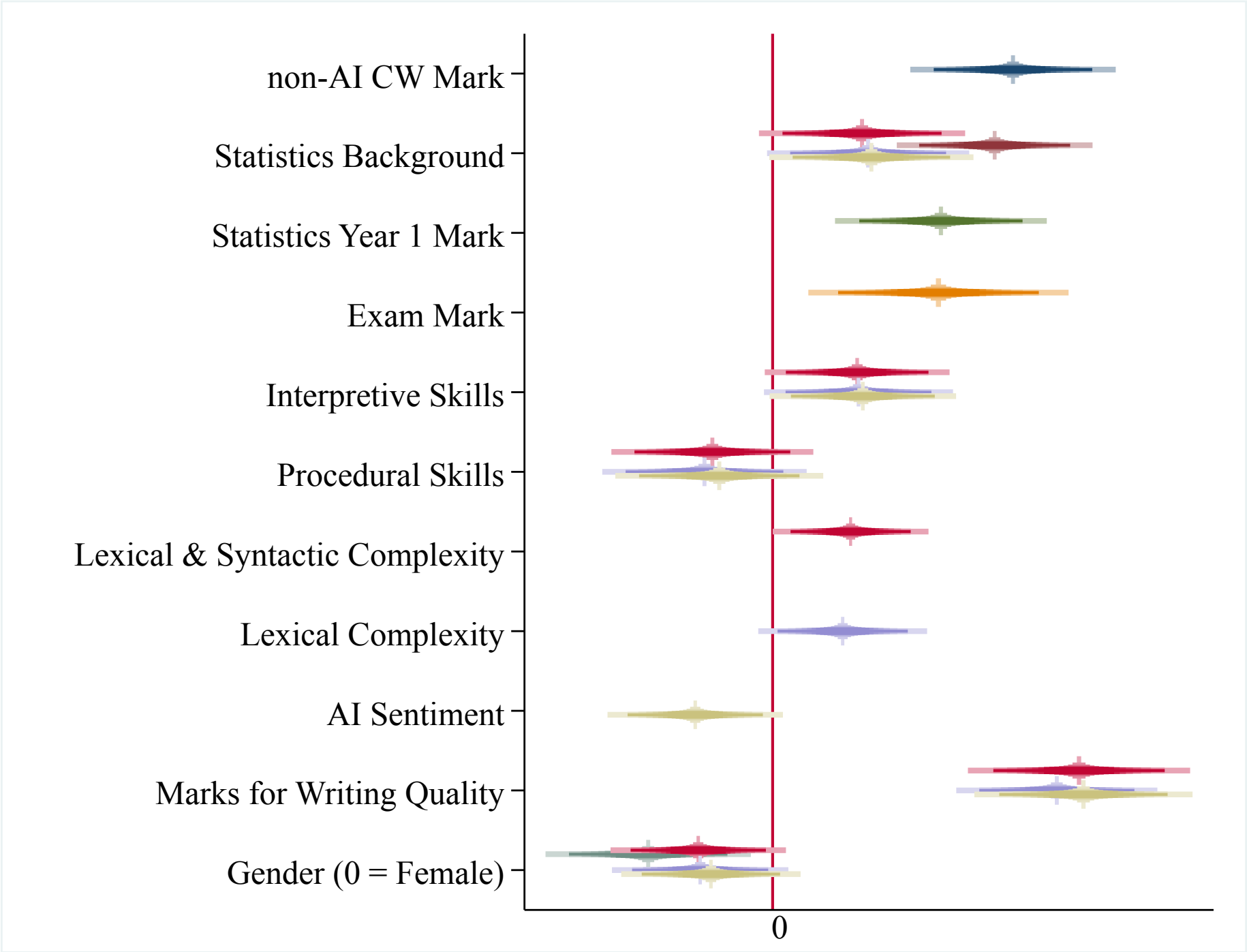

Panel B

***Notes:*** Point estimates and smoothed confidence intervals for the key predictors from OLS. The dependent variable in Panel A is *Detection,* in Panel B is the *AI mark.* Different colours represent different estimation specifications. The intensity of colours represents the different confidence levels. The visible dark colours are at the 90% confidence interval. For presentation purposes, the coefficients are rescaled by the variables' standard deviations.

**Stronger language and writing skills** enhance students' ability to detect and articulate AI hallucinations. While lexical complexity alone is a weak predictor, students who use clear sentence structures and coherent writing are more likely to identify AI errors. *Marks for Writing*, which captures writing cohesion and clarity, reflects metacognitive regulation and epistemic cognition, as students who assess logical flow, consistency, and subject-specific nuances are better at recognising AI's weaknesses. These findings align with cognitive bias and epistemic scepticism research, highlighting the role of academic skills in AI literacy and high-stakes performance. In management education, integrating AI literacy into curricula is essential for developing critical evaluation and analytical writing skills—key competencies for decision-making in AI-assisted environments. Strengthening these skills reduces automation bias, ensuring students critically assess AI-generated insights.

Interestingly, **female students** exhibit higher AI hallucination detection rates, though no difference is observed in their ability to articulate AI errors. As explored in Appendix B, writing skills—along with critical thinking (interpretive skills) and attention to detail (procedural skills)—play a key role in this gender-based difference. Our results contrast Peren Arin et. al (2023) who report that females in the UK are less likely to detect political fake news regardless of education. However, our results align with the broader literature suggesting that females tend to be more cautious toward AI, and, as we show below, AI scepticism strongly correlates with hallucination detection.

Indeed, we find that students with a more negative view of AI and its usages were more likely to detect and articulate the issues of AI hallucinations. Coefficients for AI Sentiment are negative and statistically insignificant at 10% in (2), (3) and (9). It is likely students who are cautious of AI are also more attentive to details and subject materials, leading to higher detection rates. The result is consistent with our hypothesis on epistemic vigilance and AI scepticism, suggesting a link between AI misinformation and psychological views.

### 4.3. Will students who successfully identify misinformation engage in similar tasks?

We now examine whether students choose the optional exam question on random sampling vs. random treatment after engaging with the AI response evaluation on sample selection bias vs. selection bias—concepts tied to the absence of random sampling and random treatment, respectively. Given that students receive detailed feedback and coursework solutions, we investigate whether evaluating the AI response influences their decision to attempt a similar question in the high-stakes exam. To test this, we estimate the following OLS regressions, using *Answer AI in Exam* (a binary variable indicating whether a student selects the optional question) and *Exam AI Mark* (their score on the question) as alternative outcome variables:

$$Y_i = \alpha + \beta_1 Detect_i + \beta_2 AI\ Mark_i + Predictors_i \gamma' \\ + MarkersCW_i \pi_1' + MarkersEX \pi_2' + e_{2i} \qquad (2)$$

where $i$ indicates student observation, $e_{2i}$ is the error terms. $\beta_1, \beta_2$ are the coefficients of interest, $\gamma$ is the vector of coefficients for the predictor variables, and $\pi_1, \pi_2$ are the vectors of coefficients for the markers fixed effects. Table 4 reports the results.

**Table 4.** *Will students engage in similar tasks in the exams after the coursework?*

| | *Answer in Exam* | | *Exam AI Mark* | |
|---|---|---|---|---|
| *Variables* | (1) | (2) | (3) | (4) |
| *Detect* | 0.056 | 0.068 | 0.045 | -0.252* |
| | (0.055) | (0.085) | (0.057) | (0.129) |
| *AI Mark* | | -0.004 | | |
| | | (0.027) | | |
| *Statistical Background* | | | 0.000 | 0.001 |
| | | | (0.008) | (0.013) |
| *Gender* | | | -0.045 | -0.016 |
| | | | (0.051) | (0.103) |
| Markers CW | Yes | Yes | Yes | Yes |
| Markers EX | Yes | Yes | Yes | Yes |
| Observations | 205 | 205 | 205 | 174 |
| Adjusted R-squared | 0.001 | -0.004 | -0.005 | 0.341 |

*Notes:* *p < .10; ** p < .05; *** p < .01 Robust standard errors in parenthesis. outcome for Column (1) and (2) is whether student choose the exam question related to the AI-coursework question. Outcome for (3) and (4) is the mark given for the exam question.

Our findings suggest little evidence that detecting or articulating AI hallucinations in the coursework increases the likelihood of choosing the same topic in the exam. Surprisingly, students who performed well in the coursework task scored slightly lower (by 0.252 on an average mark of 3.9) on the corresponding exam question than those who failed to detect AI errors. Additionally, we find no significant effect of prior statistical background or gender on students' decision to attempt the optional exam question. These results align with Transfer of Learning Theory, which posits that knowledge application depends on how learning experiences are structured. While prior success in coursework may increase content familiarity, it does not necessarily translate into task selection or improved performance in a higher stake setting. The lack of a confidence-boosting effect contradicts the expectation that mastery experiences would lead to greater willingness to re-engage with similar tasks.

A plausible explanation lies in the Active Learning Strategies embedded in our course. By providing detailed feedback and model solutions, we enable students—especially those who initially struggled—to revise, prepare, and bridge knowledge gaps before the exam. This aligns with self-regulated learning theory (Zimmerman, 2002) where students actively refine their understanding based on feedback rather than relying solely on prior coursework success. As a result, by the time of the exam, initial differences in AI hallucination detection no longer influence students' choices or outcomes, as all students are equally prepared for the task. These findings challenge the assumption that early success in AI-related assessments fosters greater confidence in future high-stakes settings, highlighting instead the importance of feedback-driven learning and adaptive engagement in AI literacy development.

## 5. Discussion and implications

Our findings reveal that academic performance, interpretive reasoning, writing coherence, and AI scepticism are key predictors of AI hallucination detection, while rote procedural knowledge plays a limited role. Female students exhibit higher detection rates, likely due to greater attention to detail and critical thinking, though their articulation ability remains comparable to male students. Additionally, students with negative sentiment toward AI are more likely to identify and critically evaluate AI-generated errors, suggesting a link between scepticism and epistemic vigilance. However, detecting AI hallucinations in coursework does not significantly influence students' choice or performance on a similar question in the final exam. The presence of detailed feedback and structured learning support appears to bridge initial performance gaps, allowing all students to engage with AI-related content on equal footing by the exam. These results emphasise the importance of active learning strategies, targeted feedback, and structured AI literacy integration in management education.

### 5.1. Design authentic assessment in economics management education with GenAI

Our study introduces a scalable assessment framework that integrates AI literacy into economics and management education, addressing limitations of traditional assessments in capturing critical thinking and epistemic cognition. With AI capable of generating coherent yet misleading responses (Jones & Bergen, 2024) and the shift toward online and unsupervised assessments (Lockee, 2021; Pieters, 2024), there is a growing need for authentic evaluation methods that go beyond rote knowledge recall.

We propose embedding AI hallucination detection into coursework, simulating real-world challenges where professionals must critically assess AI-generated content. This approach cultivates pragmatic judgment skills essential in business education (Larson et al., 2024; Mason et al., 2024; Moser et al., 2022) encouraging students to actively verify and validate AI-generated information rather than passively consuming it.

### 5.2. Faculty Development: Enhancing educators' skills for AI-integrated assessment

To implement AI-integrated assessment strategies effectively, faculty should develop expertise in AI literacy, epistemic cognition, and cognitive biases. This goes beyond technical familiarity with AI tools, requiring the ability to guide students in evaluating AI-generated content, identifying hallucinations, and applying structured reasoning to verify outputs. Faculty training should emphasise metacognitive scaffolding, enabling educators to help students reflect on their reasoning processes and recognise misinformation patterns. Given the risk of automation bias, faculty should also integrate exercises that foster epistemic scepticism and structured verification skills. This development becomes ever important in our society.

Institutions should support faculty development through structured training programmes, interdisciplinary collaboration, and AI-integrated pedagogy. Professional development workshops should equip educators with strategies to design assessments that require students to critically analyse AI-generated responses rather than passively accept them. Encouraging cross-disciplinary collaboration among business, education, and AI researchers will ensure best practices in AI-enhanced coursework design. Embedding AI literacy systematically across curricula—rather than treating it as an isolated skill—will better prepare students to engage critically with AI in both academic and professional settings, strengthening their analytical and decision-making competencies in management education.

### 5.3. Curriculum integration: Systematic implementation of AI literacy

Our findings highlight the importance of writing coherence, reasoning, and structured analysis in detecting AI hallucinations, reinforcing the need to embed AI literacy across the curriculum rather than treat it as a standalone topic. Courses should integrate critical thinking and epistemic scepticism into analytical reasoning, writing, and decision-making, with early focus on information verification and advanced modules incorporating AI-assisted assignments.

AI literacy should be embedded in learning objectives, ensuring students engage with AI across multiple contexts. Business and management courses should incorporate structured AI sensemaking exercises, requiring students to critically assess AI-generated content. Capstone courses should further develop these skills in real-world decision-making.

Assessments must evolve to require students to evaluate and challenge AI-generated outputs, strengthening their understanding of AI's epistemic limitations. Our study's high-stakes but low-time-pressure setting, aligned with Cognitive Load Theory, allows students to critically engage with AI. In fast-paced classrooms, where cognitive load is higher, hallucinations may go unnoticed without explicit AI-critical reasoning training.

If AI hallucination is used as a teaching tool rather than for assessment, students may engage with its outputs passively, reinforcing automation bias. Active interventions, such as structured exercises requiring students to verify AI claims and debate accuracy, are essential to fostering critical AI engagement.

### 5.4. The long-term significance of AI hallucinations

Recent research has contrasting views on the persistence of AI hallucinations. Some argue that hallucinations are an inherent limitation of large language models, particularly in complex reasoning, context sensitivity, and factual alignment (Mirzadeh et al., 2024). As such, hallucinations are not temporary flaws but rather an unavoidable byproduct of probabilistic text generation, making complete elimination unlikely. Conversely, advancements in retrieval-augmented generation (RAG), reinforcement learning from human feedback (RLHF), and knowledge-grounding techniques have shown measurable reductions in hallucination rates (Bubeck et al., 2023). As AI integrates real-time fact verification and enhanced reasoning capabilities, its reliability will steadily improve.

Given these possibilities, our action plan must account for both scenarios. If AI hallucinations become negligible, the emphasis of AI literacy and critical thinking education will shift from error detection to deeper engagement with AI's decision-making processes. This aligns with epistemic AI literacy, which stresses the importance of interrogating sources, understanding model limitations, and assessing AI-generated content beyond factual accuracy (for example see (Annapureddy et al., 2024) As AI systems incorporate more robust verification mechanisms, curricula may need to focus on interpretability and trust calibration, helping students determine when and how to rely on AI outputs rather than simply identifying errors. These skills should become part of broader digital competencies.

If hallucinations persist, however, our findings highlight the urgent need to integrate AI-critical evaluation skills into curriculum design. Pedagogical approaches must provide structured frameworks for verifying AI-generated content, integrating cross-referencing techniques and domain-specific detection strategies to differentiate accurate from misleading outputs. From an assessment perspective, persistent hallucinations reinforce the necessity for AI-enhanced assignments that require students not only to generate AI-supported analyses but also to produce meta-analyses on the validity, reliability, and limitations of AI outputs. This aligns with the growing discourse on AI-assisted assessment, where students are expected to engage critically, rather than passively, with AI-generated material.

### 5.5. Limitations and future research

Grounded in cognitive load theory, our study examines AI hallucination detection in a high-stakes assessment with ample preparation time. In time-constrained settings, such as classrooms or professional interactions, evaluating AI-generated outputs remains challenging. Future research could replicate our findings across different disciplines and contexts, using diverse participant groups. While prior exposure to and experience with AI tools are relevant predictors, we cannot capture them during our assessment setting. Instead, we proxy epistemic literacy through AI sentiment, though direct measures (via self-reported surveys, GenAI competency tasks, or prior AI interactions) would provide more valuable insights.[3] Finally, as LLM tools continue to evolve, comparing their hallucination tendencies presents a valuable avenue for future research.

[3] We thank an anonymous referee for this insightful suggestion.

## ONLINE APPENDIX A: Measures of text readability[1]

1. *Flesch-Kincaid Readability Score (Flesch and Kincaid, 1975)*

The Flesch-Kincaid Readability Score (Readability_FKR) is a measure used to determine how easy a text is to read. It is widely used in educational settings to assess the readability of textbooks and other educational materials. The score is based on two main factors: the average number of syllables per word and the average number of words per sentence.
The formula for calculating the Flesch-Kincaid Readability Score is as follows:

$$Readability_{FKR} = 0.39 \times \left(\frac{Total\ Words}{Total\ Sentences}\right) + 11.8 \times \left(\frac{Total\ Syllables}{Total\ Words}\right) - 15.59$$

Where:

• Total Words: The total number of words in the text.

• Total Sentences: The total number of sentences in the text.

• Total Syllables: The total number of syllables in the text.

2. *New Dale-Chall Readability Formula (Chall and Dale, 1995)*

The New Dale-Chall Readability Formula (Readability_CDale) is another popular method used to assess the readability of a text. This formula uses a list of 3,000 familiar words that fourth-grade students in the U.S. are expected to know. The score is calculated based on the percentage of words in the text that are not on the familiar word list, as well as the average number of words per sentence.
The formula for calculating the New Dale-Chall Readability Score is as follows:

$$\begin{aligned} Readability_{CDale} &= 0.1579 \times (Percentage\ of\ Difficult\ Words) \\ &+ 0.0496 \times \left(\frac{Total\ Words}{Total\ Sentences}\right) \\ &+ 3.6365\ (if\ Percentage\ of\ Difficult\ Words > 5\%) \end{aligned}$$

Where:

• Percentage of Difficult Words: The percentage of words in the text that are not found in the Dale-Chall familiar word list.

• Total Words: The total number of words in the text.

• Total Sentences: The total number of sentences in the text.

[1] We also experiment with Coleman-Liau Estimated Cloze Percent (ECP) (Coleman and Liau 1975), The Navy's Adaptation of Gunning's Fog Index (Kincaid, Fishburne, Rogers and Chissom 1975), Simple Measure of Gobbledygook (SMOG) (McLaughlin 1969), which yield similar results.

## ONLINE APPENDIX B: Gender, AI Hallucination, and potential mechanism

We explore the mechanism through which female students in our sample are more likely to detect AI hallucinations but not necessarily articulate the hallucination in their course. To do so, we alternatively add interaction terms of *Gender* with the three main predictors, namely, academic skills (*Statistics Background)*, writing skills (*Marks for Writing)*, and AI sentiment into Equation (1b). The results are reported in Columns (1) to (3) in Table A1. In columns (4) to (6), we replace the outcome variable by *AI mark* for the articulation of the student.

$$Detect_i = \alpha + Predictors_i\beta' + MarkersCW_i\gamma_1' + MarkersEX\gamma_2' + Interaction + e_i (1b)$$

In brief, we fail to detect any significant mechanism through which gender may have an effect on a student's ability to detect AI hallucination, apart from writing skills. As shown in Column (1) female students with better marks for their reflective paragraphs are even more likely to detect AI hallucination than their male counterparts.

***Table A1.*** OLS estimations on the potential mechanism of the gender-AI hallucination detection relation

| | *Detect* | | | *AI Mark* | | |
|---|---|---|---|---|---|---|
| *Predictors* | (1) | (2) | (3) | (4) | (5) | (6) |
| *Statistics Background [30]* | 0.013** | 0.014** | 0.018 | 0.047* | 0.048** | 0.051 |
| | (0.007) | (0.007) | (0.011) | (0.024) | (0.024) | (0.037) |
| *Interpretive Skills [15]* | 0.023* | 0.021 | 0.021 | 0.105** | 0.102** | 0.101** |
| | (0.014) | (0.014) | (0.014) | (0.050) | (0.050) | (0.049) |
| *Procedural Skills [15]* | -0.030*** | -0.029** | -0.029** | -0.041 | -0.039 | -0.040 |
| | (0.011) | (0.011) | (0.011) | (0.036) | (0.037) | (0.037) |
| *Gender (=0 if Female)* | 0.735*** | -0.061 | 0.017 | 0.947 | -0.160 | -0.213 |
| | (0.181) | (0.070) | (0.261) | (0.926) | (0.207) | (0.886) |
| *Marks for Writing [5]* | 0.267*** | 0.099*** | 0.094*** | 1.078*** | 0.839*** | 0.822*** |
| | (0.051) | (0.029) | (0.029) | (0.227) | (0.137) | (0.135) |
| *AI Sentiment* | -0.006 | -0.003 | -0.009 | -0.029 | -0.014 | -0.034* |
| | (0.006) | (0.010) | (0.006) | (0.019) | (0.029) | (0.018) |
| *Gender* x *Marks for Writing* | -0.224*** | | | -0.331 | | |
| | (0.055) | | | (0.257) | | |
| *Gender* x *AI Sentiment* | | -0.011 | | | -0.036 | |
| | | (0.012) | | | (0.036) | |
| *Gender* x *Statistics Background* | | | -0.005 | | | -0.003 |
| | | | (0.011) | | | (0.037) |
| Observations | 211 | 211 | 211 | 211 | 211 | 211 |
| Adjusted R-squared | 0.232 | 0.202 | 0.199 | 0.398 | 0.396 | 0.393 |

*Notes:* *** p<0.01, ** p<0.05, * p<0.1. Numbers in brackets [] indicate the maximum possible marks for the questions. Robust standard errors are in parentheses. The outcome variable for (1)-(3) is *Detect (*whether a student can detect AI hallucinations), and for (4)-(6) is *AI Mark,* how well the student can articulate their evaluation of the AI-generated response. Coefficients for constants and the fixed effects are omitted to save space. We include, alternatively, the interactions of Gender with the variables for academic skills, writing skills, and AI sentiment to see how the effects of the varying skills on detection and articulation are related to gender.